\documentclass[12pt,preprint]{aastex}

\shorttitle{Far-infrared dust emission in the halo of NGC~253}
\shortauthors{Kaneda et al.}

\begin{document}


\title{AKARI Detection of Far-Infrared Dust Emission in the Halo of NGC~253}


\author{Hidehiro Kaneda\altaffilmark{1}, Mitsuyoshi Yamagishi\altaffilmark{1}, Toyoaki Suzuki\altaffilmark{2}, and Takashi Onaka\altaffilmark{3}}
 \altaffiltext{1}{Graduate School of Science, Nagoya University, Chikusa-ku, Nagoya 464-8602, Japan}
 \email{kaneda@u.phys.nagoya-u.ac.jp}
 \altaffiltext{2}{Advanced Technology Center, National Astronomical Observatory of Japan, Mitaka, Tokyo 181-8588, Japan}
 \altaffiltext{3}{Graduate School of Science, The University of Tokyo, Bunkyo-ku, Tokyo 113-0033, Japan}


\begin{abstract}
We present new far-infrared (FIR) images of the edge-on starburst galaxy NGC~253 obtained with the Far-Infrared Surveyor (FIS) onboard {\it AKARI} at wavelengths of 90 $\mu$m and 140 $\mu$m. We have clearly detected FIR dust emission extended in the halo of the galaxy; there are two filamentary emission structures extending from the galactic disk up to 9 kpc in the northern and 6 kpc in the northwestern direction. From its spatial coincidence with the X-ray plasma outflow, the extended FIR emission is very likely to represent outflowing dust entrained by superwinds. The ratios of surface brightness at 90 $\mu$m to that at 140 $\mu$m suggest that the temperatures of the dust in the halo are getting higher in the regions far from the disk, implying that there exist extra dust heating sources in the halo of the galaxy.

\end{abstract}





\section{Introduction}
Starbursts are important mechanisms not only for chemical processing of elements, but also as sources of material fed into intergalactic space. Massive starburst phenomena, i.e. supernova explosions and energetic stellar winds, can drive gas from the disks of host galaxies to their halos, which are observed for many local starburst galaxies as superwind or X-ray-emitting hot plasma accelerated along the minor axes (e.g. Lehnert \& Heckman 1996). The outflowing hot plasma may be capable of entraining cold neutral dust-bearing gas to a substantial height above the galactic disk, of which mechanism is yet poorly understood. Dust grains are indeed found to spread in the vertical directions at optical wavelengths over larger scales than was thought earlier (Rossa et al. 2004; Dettmar 2005); dust grains should be strongly coupled to magnetic fields once they are charged, and thus could be trapped in the disk region. Grain destruction in the hot halo environment may also prevent such an expulsion (Ferrara et al. 1991).
%The physical relationship between the cool dust-bearing gas and hot gas probed by X-ray is still unclear.
Whether the dust can escape or not is an important question, because this would enrich the intergalactic medium with dust, which could affect observations of high-redshift objects (e.g. Heisler \& Ostriker 1988; Davies et al. 1998). The dust expulsion from a galaxy would also play an important role in galactic chemical evolution acting as a sink for heavy elements (e.g. Eales \& Edmunds 1996). However, no comprehensive view has been observationally established for the vertical distribution of dust grains in disk galaxies. This is largely owing to a lack of spatially resolved sensitive far-infrared (FIR) images of external galaxies.

NGC~253 is one of best local laboratories to investigate the vertical distribution of such extraplanar dust due to its proximity (distance: 3.5$\pm$0.2 Mpc; Rekola et al. 2005), nearly edge-on orientation (inclination angle: 78.5 $^{\circ}$; Pence 1980), and strong nuclear starburst (Dudley \& Wynn-Williams 1999). The halo of NGC~253 has been studied at various wavelengths (e.g. X-ray: Dahlem et al. 1998; Pietsch et al. 2000; Strickland et al. 2002; Bauer et al. 2008; UV: Hoopes, et al. 2005; H$\alpha$: Hoopes et al. 1996; HI: Boomsma et al. 2005). FIR observations are crucial to look for evidence of the dust outflow since dust emission typically peaks in the FIR. The nuclear source is, however, extremely strong in the FIR, which severely affects the detection of faint extended emission outside the disk. For the {\it IRAS} observation of NGC~253, Rice (1993) and Alton et al. (1998) concluded that instrumental effects masked low-level FIR emission outside the disk. With {\it ISO}, Radovich et al. (2001) and Melo et al. (2002) studied the spatial distribution of the FIR emission in NGC~253, where they detected FIR excess emission along the minor axis. Its spatial structures were, however, somewhat controversial mainly due to the poor spatial resolution, and the central peak of the galaxy was saturated (Melo et al. 2002). {\it Spitzer} observed NGC~253, the FIR distribution of which is not yet reported in any paper; we find that the MIPS archival data show strong artifacts and severe signal saturation in the 70 $\mu$m and the 160 $\mu$m band, respectively. From submillimeter observations of the nuclear region of NGC~253, Alton et al. (1999) found evidence for a dust outflow along the minor axis at 450 $\mu$m on a small scale $\leq$ $45''$.  Tacconi-Garman et al. (2005) and Sugai et al. (2003) detected extraplanar polycyclic aromatic hydrocarbon (PAH) and H$_2$ emissions, respectively, both on small scales of $\sim$ $10''$ along the minor axis.   

We performed FIR observations of NGC~253 with the Far-Infrared Surveyor (FIS; Kawada et al. 2007) on board {\it AKARI} (Murakami et al. 2007).
The fundamental advantage the {\it AKARI}/FIS offers over any other previous or currently existing instruments is a combination of its high saturation limits, high sensitivity, and relatively high spatial resolution, which is essential to detect faint FIR dust emission extending to the halo of an edge-on starburst galaxy. Hence the {\it AKARI} FIR images of NGC~253 are expected to be much less affected by very strong nuclear emission of the galaxy than those previously obtained with {\it IRAS} and {\it ISO}.  
%We below present new FIR images of NGC~253 obtained with the FIS.  

\section{Observations}
We observed NGC~253 with the {\it AKARI}/FIS in part of the {\it AKARI} mission program ``ISM in our Galaxy and Nearby Galaxies'' (ISMGN; PI: H. K.). The observation log is listed in Table 1. The FIS observation was performed in the special fast reset mode called CDS (Correlated Double Sampling) mode in order to avoid signal saturation near the nuclear region of the galaxy. Thus fluxes were later cross-calibrated with the other ordinary integration modes by using the internal calibration lamp of the FIS. A $10'\times 30'$ region was covered by using one of the FIS observation modes, FIS01, a 2-round-trip slow scan with a scan speed of $15''$ s$^{-1}$ and a cross-scan step of $240''$ between the first and the second round trip. Due to limited visibility to {\it AKARI}, NGC~253 could be observed with the FIS only once, and therefore we could not cover the whole disk of the galaxy; the scan passed the central part of the galaxy and the scan direction is nearly parallel to the minor axis. 

The FIR images were processed from the FIS Time Series Data (TSD) by using the {\it AKARI} official pipeline developed by the {\it AKARI} data reduction team (Verdugo, E. et al. 2007). The images were further cleaned by removing the after-effects of cosmic-ray hits and the latency of the illumination of the internal calibration lamps with the algorithm developed in Suzuki (2007) and Suzuki et al. (2009), where the correction for the variations of detector responsivity was also performed. Any residual non-uniformity of the responsivity within the detector arrays was carefully removed by assuming spatially flat blank skies near both ends of the scan, which are far enough from the galactic disk. 

With the FIS, we obtained N60 (centered at a wavelength of 65 $\mu$m with the effective band width of 22 $\mu$m), WIDE-S (90 $\mu$m with 38 $\mu$m), WIDE-L (140 $\mu$m with 52 $\mu$m), and N160 (160 $\mu$m with 34 $\mu$m) band images. The 2 narrow-bands, N60 and N160, however, suffer ghost image problems originating from the very bright central source (Kawada et al. 2007). They also have relatively low S/Ns for detecting halo emission. Since our purpose is to firmly detect dust in the halo and estimate its temperature and mass, but not to discuss the spectral energy distribution of the dust emission, we below concentrate on the results of the 2 wide-bands, WIDE-S and WIDE-L, for which 5$\sigma$ detection limits in the above observation mode are 0.7 and 1.2 Jy, respectively. Nevertheless, the nuclear source was so extremely bright that it may have been somewhat affected by instrumental effects. Hence, in this paper, we do not discuss any detailed spatial structure of FIR emission near the nuclear region. 

\section{Results}
Figure 1 shows the WIDE-S (90 $\mu$m) and WIDE-L (140 $\mu$m) images of NGC~253 obtained with the {\it AKARI}/FIS. Note that the spatial coverage of the WIDE-S image is relatively small and shifted as compared to that of the WIDE-L image due to the difference in the field-of-view between the corresponding detector arrays. The background sky levels are estimated by averaging surface brightness of nearby blank sky in 4 regions with the area of $95''\times 95''$ for each, where 2 in the south and 2 in the north with respect to the disk are selected to avoid any overlap with faint extended emission from the galaxy, and then subtracted from the images. Background fluctuation levels are also estimated from these regions, which are approximately 0.6 MJy/str for both band images. We have applied boxcar smoothing with the width of $57''$ to both band images in order to bring out low surface brightness emission in the halo. The resultant spatial resolution is still significantly higher than the actual resolution ($\sim 180''$) of the {\it ISO} map of NGC~253 (Melo et al. 2002). 

As can be seen in the figure, we clearly detect FIR dust emission extended in the halo of the galaxy at both wavelengths. In the 90 $\mu$m image, there are two filamentary components extending from the galactic disk up to $9'$ or 9 kpc and $6'$ or 6 kpc by adopting the distance of 3.5 Mpc (Rekola et al. 2005), in the northern and northwestern directions; the former is clearly seen also in the 140 $\mu$m image. The {\it ISO} image showed excess emission in the northern direction, probably corresponding to the former component (Radovich et al. 2001). However we do not see any star-like patterns centered at the nucleus as seen in the {\it IRAS} and {\it ISO} images (Radovich et al. 2001). Instead, dust halo emission is highly suppressed in the southern part of the galaxy, and thus there is a significant difference in the spatial distribution of dust emission between the northern and southern halo regions.  
 
In table 2, we derive the flux densities of the center, halo 1, and halo 2 regions, the positions of which are given in Fig.1, by integrating the surface brightness within a circular aperture diameter of $4'$ in each band image of Fig.1. Since the aperture size is sufficiently large, we do not consider aperture corrections. Instead we consider the flux uncertainties including both systematic effects associated with the FIR detectors and absolute calibration uncertainties, which are estimated to be no more than 20 \% for WIDE-S, and 30 \% for WIDE-L (Kawada et al. 2007). The values obtained with the {\it AKARI}/FIS in table 2 show overall agreements with the {\it IRAS} and {\it ISO} results (Radovich et al. 2001); as for the flux densities in the center region, our results show slightly higher values than those with {\it IRAS} and {\it ISO}, even though the latter fluxes were obtained with twice larger aperture size ($8'$). For {\it ISO}, the discrepancy will be probably due to the effect of the signal saturation (Melo et al. 2002). 

For the emissivity power-law index of $\beta=1$, the ratios of the WIDE-S to the WIDE-L flux density correspond to the temperatures of FIR dust of 31 K, 20 K, and 23 K for the center, the halo 1, and the halo 2 region, respectively, which are averaged over the circular aperture of the diameter of $4'$. We calculate dust mass by using the equation (e.g. Hildebrand 1983):
\begin{equation}
M_d=\frac{4a\rho D^2}{3}\frac{F_{\nu}}{Q_{\nu}B_{\nu}(T)},
\end{equation}
where $M_d$, $D$, $a$, and $\rho$ are the dust mass, the galaxy distance, the average grain radius, and the specific dust mass density, respectively. $F_{\nu}$, $Q_{\nu}$, and $B_{\nu}(T)$ are the observed flux density, the grain emissivity, and the value of the Planck function at the frequency of $\nu$ and the dust temperature of $T$. We adopt the grain emissivity factor given by Hildebrand (1983), the average grain radius of 0.1 $\mu$m, and the specific dust mass density of 3 g cm$^{-3}$. We use the 90 $\mu$m flux densities in Table 2 and the above dust temperatures. The dust masses thus derived from the halo 1 and halo 2 regions are $1.1\times 10^6$ M$_{\odot}$ and $9.1\times 10^5$ M$_{\odot}$, respectively. We should note that the FIS is insensitive to colder dust emitting predominantly at submillimeter wavelengths (e.g., Alton et al. 1999). Hence the above dust masses are considered to be lower limits to the real dust mass contained in these regions. These dust masses are, however, still unlikely to make a significant contribution to the total dust mass, $3.5\times 10^7$ M$_{\odot}$, integrated over the whole galaxy of NGC~253 (Radovich et al. 2001).  

Figure 2 gives the {\it ROSAT}/PSPC X-ray (0.1--2.4 keV) contour map of NGC~253 overlaid on the WIDE-S image, where a spatial correlation between the X-ray and FIR images can be recognized. H$\alpha$ emission in the halo of NGC~253 is very similar in morphology to the diffuse X-ray emission (Strickland et al. 2002). More recent {\it XMM-Newton} X-ray data revealed very similar distribution in the 0.5--1.0 keV band resembling a horn structure (Bauer et al. 2008). The X-ray halo emission is neither spatially nor spectrally uniform; the southern halo is softer than the northern halo (Bauer et al. 2008). From its spatial coincidence with the X-ray plasma outflow, the extended FIR emission is very likely to represent outflowing dust entrained by superwinds. The PAH emission extends toward the halo 1 direction in a 0.1 kpc scale from the disk (Tacconi-Garman et al. 2005), while several galaxies show extraplanar PAHs extended up to several kpc scales (e.g. Irwin \& Madden 2006; Whaley et al. 2009); the PAHs might have been similarly ejected from the disk. The extraplanar HI is found only in the northern half of the galaxy (the bottom panel of Fig.2; Boomsma et al. 2005), where its mass is estimated to be $8\times 10^7$ M$_{\odot}$, giving a dust-to-gas mass ratio $\gtrsim$ 1/40 in the northern halo, somewhat higher than the ratio for the whole galaxy $\sim$1/70 (Radovich et al. 2001). From the synchrotron radio measurements of the halo of NGC~253, Heesen et al. (2009) showed that the cosmic-ray transport is convective and more efficient in the northern halo, while it is diffusive in the southern halo, which can explain the different amounts of extraplanar HI, H$\alpha$, dust, and soft X-ray emission between the northern and the southern halo.

For the sputtering time of dust grains with the size of 0.1 $\mu$m embedded in the hot plasma, the {\it XMM-Newton} X-ray observations indicated 4--30 Myr from the plasma density between $2.5\times 10^{-2}$ cm$^{-3}$ in the outflow close to the center and $3.2\times 10^{-3}$ cm$^{-3}$ out in the northwestern halo (Bauer et al. 2008). Hence, if we assume that the observed dust is homogeneously mixed in the hot plasma, an averaged dust outflow velocity must be in a range of 300--2000 km s$^{-1}$ in order for the dust grains to reach their present locations, 9 kpc above the disk (halo 1), during the sputtering destruction timescale. Since the escape velocity is estimated to be 280 km s$^{-1}$ from the galaxy rotation speed of NGC~253 (Heesen et al. 2009), the dust would then be escaping from the gravitational potential of NGC~253. However we should note that the dust entrained in the superwind is likely in dense clumps and thus may not all get sputtered away at the same rate as in the above homogeneous scenario (e.g. Tacconi-Garman et al. 2005).  

There appears to be a difference between WIDE-S (90 $\mu$m) and WIDE-L (140 $\mu$m) in spatial structure of the dust halo emission (Fig.1), which suggests that the dust temperature is not uniform in the halo. In Fig.3, we calculate the ratios of surface brightness at 90 $\mu$m to that at 140 $\mu$m. We adjust the peak positions of the images precisely by interpolating data between pixels so that they can coincide within the accuracy of $1''$ before the calculation. In dividing the images, we use the heavily-smoothed images of Fig.1, whereby we reduce the effects of differences in the beam size between the WIDE-S (FWHM: $39''$) and WIDE-L ($58''$) bands (Kawada et al. 2007). The area in the WIDE-L image with brightness levels lower than 4.0 MJy/str that is about 7 times larger than the above background fluctuation is masked. As a result, the ratio map in Fig.3 reveals that the temperatures of the dust in the halo are significantly lower than those in the nuclear and disk regions. More interestingly, the ratio map suggests that the temperatures of the dust in the halo 1 and halo 2 regions are getting higher in the regions far away from the disk. By assuming $\beta=1$, again, the maximal ratios correspond approximately to the dust temperatures of 48 K for the center region, 33 K for the halo 1, and 31 K for the halo 2 regions, while the dust temperatures elsewhere in the halo are below 20 K. Uncertainties in the temperature for the halo 1 and halo 2 regions originating from the errors of surface brightness are about 2 K, and thus the increase in the temperature is significant.     

The excess dust in the halo is more or less heated by stellar UV photons propagating up from the disk of NGC~253. Figure 3 implies that there exist extra dust heating sources in the halo of the galaxy in addition to the stellar photons coming from the disk. As seen in Fig.2, the high-temperature region in the halo 1 of Fig.1 spatially corresponds to the border between the HI plume (Boomsma et al. 2005) and the X-ray halo emission (e.g. Bauer et al. 2008), which may reflect that the dust in the halo 1 is heated by shock through interaction between the X-ray superwind and pre-existing HI halo clouds as proposed by the model (a) of Strickland et al. (2002).
%pre-existing halo clouds are shock-heated by the superwind, as proposed by the model (a) of Strickland et al. (2002). If this scenario is correct, our result implies that the cloud is not primordial, since it contains dust. 
In contrast, no corresponding HI structures have been found near the high-temperature region in the halo 2 (Fig.2), and thus we may have to seek another heating mechanism for the dust in this region, where there is enhancement in the X-ray emission; one possibility is that the dust may be heated up to higher temperatures through collisions with plasma electrons (Dwek 1986). 

\section{Conclusions}
With the {\it AKARI}/FIS, we clearly detect FIR dust emission extended in the halo of the edge-on starburst galaxy NGC~253. We find that there are two filamentary emission structures extending from the galactic disk up to 9 kpc in the northern and 6 kpc in the northwestern direction. From its spatial coincidence with the X-ray plasma outflow, the extended FIR emission is very likely to represent outflowing dust entrained by superwinds. The temperatures of the dust in the halo are getting higher in the regions far away from the disk, implying that there exist extra dust heating sources in the halo of the galaxy other than stellar UV photons from the disk.

\acknowledgments
We thank all the members of the {\it AKARI} projects, particularly those belonging to the working group for the ISMGN mission program. We would also express many thanks to the anonymous referee for giving us useful comments. {\it AKARI} is a JAXA project with the participation of ESA. This research is supported by the Grants-in-Aid for the scientific research No. 19740114 and the Nagoya University Global COE Program, "Quest for Fundamental Principles in the Universe: from Particles to the Solar System and the Cosmos", both from the Ministry of Education, Culture, Sports, Science and Technology of Japan.

\begin{figure}
\epsscale{.6}
\plotone{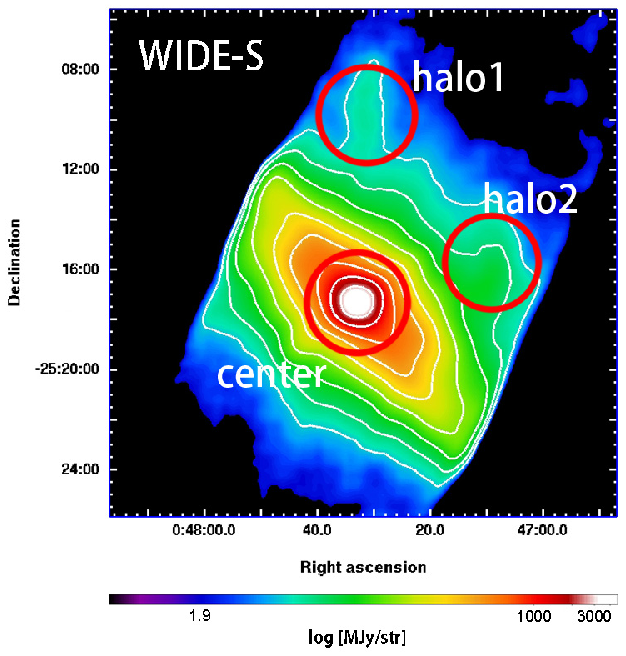}
\plotone{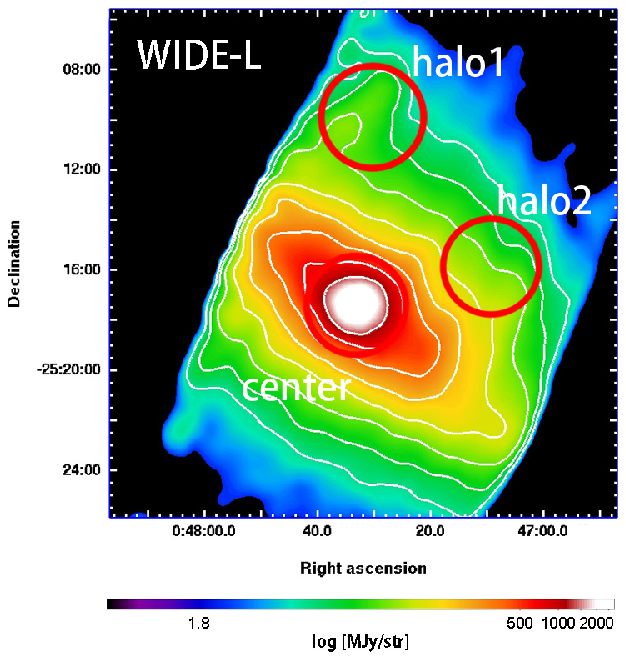}
\caption{FIR images of NGC~253 obtained with the {\it AKARI}/FIS in the WIDE-S and WIDE-L bands at the central wavelengths of 90 $\mu$m and 140 $\mu$m, respectively. The contours are drawn at surface brightness levels of 10, 20, 40, 80, 160, 320, 640, 1280, and 2560 MJy/str (the last one is only for WIDE-S). }
\end{figure}

%\clearpage

\begin{figure}
\epsscale{.7}
\plotone{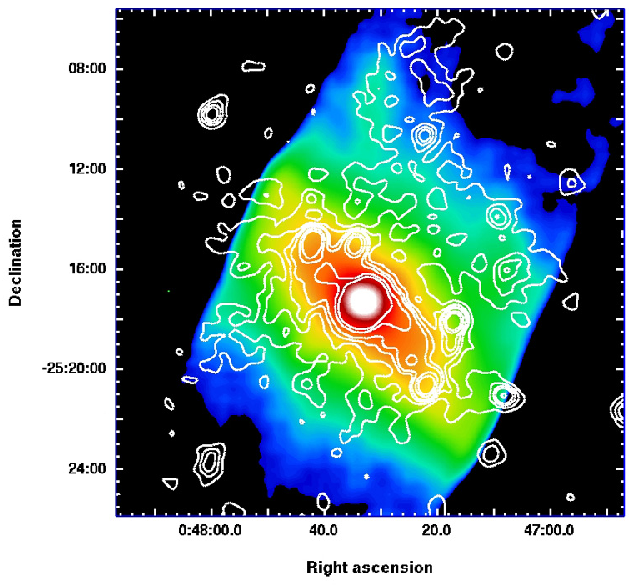}
\plotone{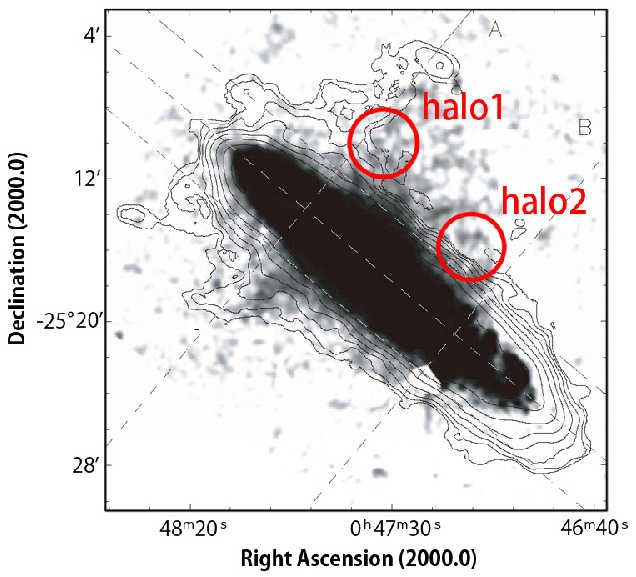}
\caption{{\it Top}: {\it ROSAT}/PSPC X-ray (0.1--2.4 keV) contour map of NGC~253 overlaid on the WIDE-S (90 $\mu$m) image with the same color levels as in Fig.1. The contours are drawn at surface brightness levels of 1.6, 3.0, 4.4, 5.8, 9.3, and 12.8$\times 10^{-3}$ counts s$^{-1}$ arcmin$^{-2}$. {\it Bottom}: HI contours overlaid on an H$\alpha$ image, same as Fig.3 of Boomsma et al. (2005), but with the circular apertures in Fig.1 added.}
\end{figure}

\begin{figure}
\epsscale{.7}
\plotone{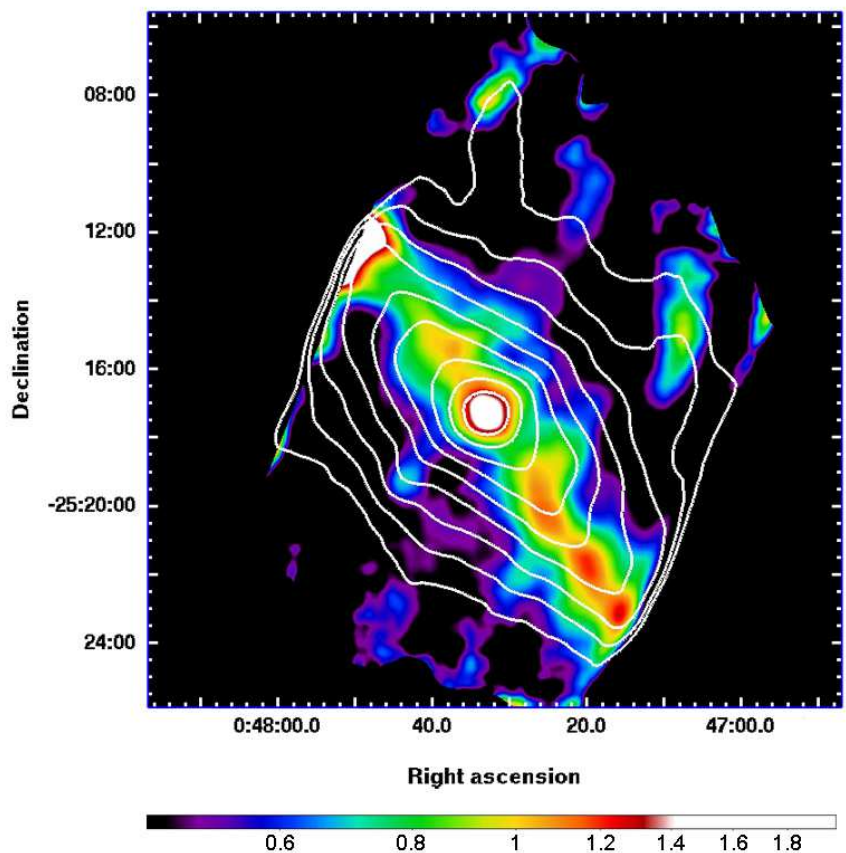}
\caption{WIDE-S (90 $\mu$m)/WIDE-L(140 $\mu$m) ratio map of NGC~253 overlaid on the WIDE-S contour map with the same contour levels as in Fig.1.}
\end{figure}

%\clearpage

\begin{deluxetable}{crrrr}
\tabletypesize{\scriptsize}
%\rotate
\tablecaption{Observation log}
\tablewidth{0pt}
\tablehead{
\colhead{Name} & \colhead{R.A. (J2000)} & \colhead{Decl. (J2000)} & \colhead{Observation ID} & \colhead{Date} }
\startdata
NGC~253 & 0 47 33.1 & $-$25 17 17.5 & 1402148 & 2007 Jun 21 \\
\enddata

Units of right ascension are hours, minutes, and seconds, and units of declination are degrees, arcminutes, and arcseconds.
%% Text for table notes should follow after the \enddata but before
%% the \end{deluxetable}. Make sure there is at least one \tablenotemark
%% in the table for each \tablenotetext.
\end{deluxetable}

\begin{deluxetable}{lrrr}
\tabletypesize{\scriptsize}
%\rotate
\tablecaption{Flux densities of the nuclear and halo regions of NGC~253}
\tablewidth{0pt}
\tablehead{
\colhead{} & Center\tablenotemark{a} & Halo 1 & Halo 2\\
 & (Jy) & (Jy) & (Jy)
}
\startdata
{\it AKARI} FIS WIDE-S 90 $\mu$m\tablenotemark{b} & 1200$\pm$240 & 9.5$\pm$1.9 & 18$\pm$3.6 \\
{\it AKARI} FIS WIDE-L 140 $\mu$m & 1300$\pm$390 & 26$\pm$7.8 & 38$\pm$12 \\
{\it IRAS} 100 $\mu$m\tablenotemark{c} & 949 & 25.5 & \\
{\it ISO} 180 $\mu$m & 607 & 28.3 & \\
\enddata
\tablenotetext{a}{The positions of the regions are indicated in Fig.1.}
\tablenotetext{b}{The {\it AKARI} flux density errors include both systematic effects associated with the FIR detectors and absolute calibration uncertainties.}
\tablenotetext{c}{The {\it IRAS} and {\it ISO} flux densities are taken from Radovich et al. (2001).}

\end{deluxetable}

\end{document}